\documentclass[twocolumn]{aastex62}

\usepackage{multirow}

\graphicspath{{./}{figures/}}

\received{March 5, 2019}
\revised{June 21, 2019}
\accepted{July 2, 2019}
\published{August 21, 2019}

\shorttitle{}
\shortauthors{Oklop{\v{c}}i{\'c}, A.}

\begin{document}

\title{Helium Absorption at 1083 nm from Extended Exoplanet Atmospheres: Dependence on Stellar Radiation}

\email{antonija.oklopcic@cfa.harvard.edu}

\author[0000-0002-9584-6476]{Antonija Oklop{\v{c}}i{\'c}}\altaffiliation{NHFP Sagan Fellow}
\affil{Center for Astrophysics $\vert$ Harvard $\&$ Smithsonian 
60 Garden Street, MS-16, Cambridge, MA 02138, USA}

\begin{abstract}

Strong absorption signatures in the helium line at 1083 nm have recently been detected in transmission spectra of several close-in exoplanets. This absorption line originates from neutral helium atoms in an excited, metastable 2$^3$S state. The population of helium atoms in this excited state is governed by the spectral shape and intensity of the incident stellar radiation field. We investigate what kind of stellar environments are most favorable for populating the metastable helium state in extended planetary atmospheres. Our results suggest that planets orbiting at close separations from late-type stars---particularly K stars---are the most promising candidates for transit absorption signals at 1083 nm. This result is supported by observations, as all four exoplanets with currently reported helium detections orbit K-type stars. In general, conditions for exciting helium atoms become more favorable at closer orbital separations, and around stars with higher levels of extreme-ultraviolet (EUV) flux, which ionizes the helium ground state, and lower levels of mid-ultraviolet (mid-UV) flux, which ionizes the helium metastable state.

\end{abstract}


\section{Introduction} 
\label{sec:intro}

Atmospheric escape and mass loss can significantly influence the evolution of planetary atmospheres, and consequently, demographics of exoplanetary systems \citep[for a recent review see][and references therein]{Owen2018}. Until recently, direct evidence of atmospheric escape relied on observations of only a few exoplanets, mostly in the ultraviolet (UV) wavelength range. Measured transit depths in the hydrogen Lyman-$\alpha$ (Ly$\alpha$) line for hot Jupiters HD~209458b and HD~189733b, and warm Neptunes GJ~436b and GJ~3470b are many times larger than their broadband optical transit depths, suggesting the existence of extended hydrogen envelopes around these planets, formed by the escaping material \citep[e.g.][]{Vidal-Madjar2003, Lecavelier2010, Ehrenreich2015, Bourrier2018}.

The sample of escaping exoplanet atmospheres observed at UV wavelengths remains small, mostly due to observational challenges. Ly$\alpha$ observations are made difficult by (a) the interstellar medium (ISM) causing strong extinction at these wavelengths and (b) the Ly$\alpha$ emission from hydrogen in the Earth's upper atmosphere (geocorona). Jointly, they make the core of the Ly$\alpha$ line (roughly $\pm 30$~km~s$^{-1}$ around the line center) practically unobservable \citep[e.g.,][]{Ehrenreich2015}. The only information available to us comes from the broad wings of the line, which probe the high-velocity tail of the escaping material, typically with blueshifts of $\sim 50-100$~km~s$^{-1}$. Physical interpretation and origin of this highly blueshifted absorption signal is still an open question, even though several accelerating mechanisms have been proposed, including stellar radiation pressure and charge exchange between the escaped material and the stellar wind \citep[e.g.,][]{Vidal-Madjar2003, Holmstrom2008, TremblinChiang2013, Bourrier2013b, Schneiter2016, Khodachenko2017, McCann2019, Debrecht2019}. Our view of gas dynamics at smaller relative velocities and in regions closer to the planet itself---where the planetary wind is being launched---remains obscured in Ly$\alpha$ observations.

\begin{deluxetable*}{ccccc}
\tablecaption{Exoplanets with Detected Helium Absorption at 1083 nm.}
\tablewidth{0pt}
\tablehead{
\colhead{Name} &
\colhead{1083 nm Transit Depth} &
\colhead{Host Star} &
\colhead{Semimajor Axis} &
\colhead{EUV Flux Density\tablenotemark{a}}\\
\colhead{} &\colhead{} & \colhead{Spectral Type}  & \colhead{[AU]} & \colhead{[erg cm$^{-2}$ s$^{-1}$ \AA$^{-1}$]} 
}
\startdata
  HD~189733b    & $0.88\%\pm 0.04\%$\tablenotemark{b }  &  K2\tablenotemark{c }   & 0.031\tablenotemark{d }  & 23.44   \\
  HAT-P-11b & $1.08\%\pm 0.05\%$\tablenotemark{e }    &  K4\tablenotemark{f }   & 0.052\tablenotemark{f }  & 4.12   \\
  WASP-69b  & $3.59\%\pm 0.19\%$\tablenotemark{g }   &  K5\tablenotemark{h } &   0.045\tablenotemark{h } &  4.08  \\
  WASP-107b & $5.54\%\pm 0.27\%$\tablenotemark{i }  &  K6\tablenotemark{j} &  0.055\tablenotemark{j }  &  5.19  \\
\enddata
\tablenotetext{}{$^a$Average flux density over the wavelength range 100-504 \AA, at the orbital distance of the planet, from \citet{Nortmann2018};  $^b $\citet{Salz2018}; $^c $\citet{Gray2003}; $^d $ \citet{DiGloria2015}; $^e $\citet{Allart2018}; $^f $\citet{Bakos2010};  $^g $\citet{Nortmann2018}; $^h $\citet{Anderson2014}; $^i $\citet{Allart2019}; $^j $\citet{Anderson2017}.}
\label{tab:host_stars}
\end{deluxetable*}

\begin{deluxetable*}{ccccc}
\tablecaption{Exoplanets with Reported Nondetection of Helium Absorption at 1083 nm.}
\tablewidth{0pt}
\tablehead{
\colhead{Name} &
\colhead{Limit on 1083 nm Transit Depth} &
\colhead{Host Star Spectral Type} &
\colhead{Semimajor Axis}\\
\colhead{} & \colhead{}& \colhead{}  & \colhead{[AU]}
}
\startdata
WASP-12b  & $59 \pm 143$~\mbox{ppm over a 70 \AA-wide bin}\tablenotemark{a}  &  G0   &  0.023  \\
HD 209458b &   $<0.84\%$ ($90\%$ confidence)\tablenotemark{b}  &  G0  & 0.047 \\
KELT-9b &    $<0.33\%$ ($90\%$ confidence)\tablenotemark{b}  &  A0 &    0.035\\
GJ 436b  & $<0.41\%$ ($90\%$ confidence)\tablenotemark{b}  &  M3.5 &  0.029    \\
GJ 1214b  & $3.8\% \pm 4.3\%$\tablenotemark{c}  &  M4.5 &  0.014 \\
\enddata
\tablenotetext{}{$^a$\citet{KreidbergOklopcic2018}; $^b$\citet{Nortmann2018}; $^c$\citet{Crossfield2019}.}
\label{tab:host_stars_nondetection}
\end{deluxetable*}

Recent theoretical modeling of the low-density and high-temperature environments of upper planetary atmospheres (thermospheres) predicted that strong transit signals in the helium line at 1083 nm could be observed for some exoplanets \citep{OklopcicHirata2018}. The first evidence of helium absorption in an exoplanet came from the \textit{Hubble Space Telescope/Wide Field Camera 3} (HST/WFC3) observations of WASP-107b by \citet{Spake2018}, showing excess absorption in the wavelength channel containing the helium line at 1083 nm. Since then, spectrally resolved helium absorption has been detected in ground-based observations of HAT-P-11b \citep{Allart2018}, WASP-69b \citep{Nortmann2018}, HD~189733b \citep{Salz2018}, and WASP-107b \citep{Allart2019}. \citet{Mansfield2018} also reported evidence of excess helium absorption in HAT-P-11b seen in the HST/WFC3 data. 

With the growing number of detections, the helium line at 1083 nm offers an excellent opportunity to study the extended upper layers of planetary atmospheres and the physics of atmospheric escape in large samples of exoplanets. As a probe of extended atmospheres, this line has two major advantages over Ly$\alpha$: (1) it is not heavily affected by the ISM or geocorona, and (2) it can be observed from the ground using high-resolution spectrographs on a number of medium-size and large telescopes. Observations of exoplanet atmospheres so far have shown that the 1083~nm line probes the altitudes of up to a few planetary radii, typically around and below the planet's Roche radius. This atmospheric region is particularly interesting for studying the physical processes responsible for generating planetary outflows.

The absorption line at 1083 nm originates from neutral helium atoms in an excited 2$^3$S (triplet) state, which is radiatively decoupled from the ground (singlet) state, and hence metastable. The population level of the metastable helium state, and therefore the strength of the absorption signal at 1083 nm, depends not only on the properties of the planetary atmosphere, but also on the intensity and spectral shape of incident stellar radiation field. Interestingly, all four exoplanets in which helium has been detected so far are hosted by fairly active K-type stars, at orbital distances of 0.031-0.055 AU (see \autoref{tab:host_stars}). In \autoref{tab:host_stars_nondetection}, we list exoplanets whose transits have been observed at the wavelength of the helium line without detecting an absorption signal from a planetary atmosphere. Planets with reported nondetections are hosted by stars of spectral types A, G, and M.

The goal of this paper is to investigate how differences in the spectral energy distribution (SED) of the host star affect the population level of the excited metastable helium in upper planetary atmospheres and, consequently, the expected absorption signal at 1083 nm. We use the results of our investigation to determine what types of stellar environments are most favorable for producing strong absorption signatures in the helium line at 1083 nm.

\section{Methods}
\label{sec:methods}

\subsection{Model of Escaping Exoplanet Atmospheres}

We model the abundance of excited metastable helium in the escaping exoplanet atmosphere using methods very similar to those described in \citet{OklopcicHirata2018}. For completeness, here we give a brief overview of the theoretical model.

We assume a spherically symmetric and radially expanding planetary atmosphere, whose radial velocity ($v$) and density ($\rho$) profiles, as functions of altitude/radius ($r$), can be described by an isothermal Parker wind \citep{Parker1958,Lamers1999}\footnote{The velocity and density solutions of the isothermal Parker wind problem can also be found analytically using the Lambert $W$ function \citep{Cranmer2004}.}:
\begin{eqnarray}
\frac{v(r)}{v_s}\exp{\left[-\frac{v^2(r)}{2v_s^2} \right]} &=& \left(\frac{r_s}{r} \right)^2\exp{\left(-\frac{2r_s}{r} +\frac{3}{2}\right)}\\
\label{eq:parker_velocity}
\frac{\rho(r)}{\rho_s} &=& \exp{\left[ \frac{2r_s}{r}-\frac{3}{2}-\frac{v^2(r)}{2v_s^2} \right]}.
\label{eq:parker_density}
\end{eqnarray}
The Parker wind model describes a radial outflow that starts with low velocities close to the planet, gradually accelerates and becomes supersonic at the radius (altitude) called of the sonic point \textbf{($r_s$)}, which is given by
\begin{equation}
r_s = \frac{GM_\mathrm{pl}}{2v_s^2},
\end{equation}
where M$_\mathrm{pl}$ is planet's mass, $G$ is the gravitational constant, and $v_s$ is the sound speed in gas with temperature $T$ and mean molecular weight $\mu m_H$ (with $k$ as the Boltzmann constant):
\begin{equation}
v_s = \sqrt{\frac{kT}{\mu m_H}}.
\end{equation}
The gas density at the sonic point ($\rho_s$) can be calculated from the mass conservation equation
\begin{equation}
\dot{M} = 4\pi r^2 \rho(r)v(r),
\label{eq:mass_cont}
\end{equation}
assuming the value of the total mass-loss rate $\dot{M}$.

Our model atmospheres are assumed to be composed of atomic hydrogen and helium in 9:1 number ratio. The initial value of the molecular weight of the atmosphere is set to 0.9~m$_H$, but is then iteratively adjusted to the altitude-weighted mean value, taking into account the population of free electrons resulting from hydrogen ionization. In addition to atmospheric composition, the main free parameters of the model are the temperature of the thermosphere and the total mass loss rate.

Assuming a steady-state (i.e. time independent) planetary wind, we calculate level populations of hydrogen and helium atoms for a gas streamline along the planet's terminator. At low densities typical of planetary thermospheres, atoms are not in local thermodynamic equilibrium, and hence we need to explicitly model the processes that affect level populations, such as photoionization, recombination, collisional transitions, and radiative decay. 

We numerically solve Equation 13 from \citet{OklopcicHirata2018} to obtain the fraction of hydrogen in the neutral/ionized state and the density of free electrons. Then, we calculate the radial distribution of helium atoms in the singlet (ground) and triplet (excited) state by solving the following set of equations: 
\begin{eqnarray}
\nonumber v \frac{\partial f_1}{\partial r} &=& (1-f_1-f_3)n_e\alpha_1 + f_3 A_{31} - f_1\Phi_1e^{-\tau_1} \\
 &-& f_1n_e q_{13a} + f_3n_e q_{31a} + f_3 n_e q_{31b} + f_3 n_{\mathrm{H}^0}Q_{31},\\
\nonumber v  \frac{\partial f_3}{\partial r} &=& (1-f_1-f_3)n_e\alpha_3 - f_3 A_{31} - f_3\Phi_3e^{-\tau_3}+f_1n_e q_{13a} \\
&-& f_3n_e q_{31a}- f_3 n_e q_{31b} -f_3 n_{\mathrm{H}^0}Q_{31} \ \mbox{.}
\label{eq:triplet}
\end{eqnarray}
Here $f_1$ and $f_3$ mark the fractions of helium in the singlet and triplet state (relative to all helium atoms/ions), respectively, $v$ is the radial gas velocity, $n_e$ is the number density of electrons, and $\alpha$ and $\Phi$ are the recombination and photoionization rate coefficients, with optical depth $\tau$ calculated using flux-averaged cross sections. All these quantities are functions of altitude. Symbols $q$ and $Q$ mark various collision rate coefficients, while $A_{31}$ is the triplet-to-singlet radiative decay rate. More information on this calculation, including the references for the rate coefficients, can be found in \citet[]{OklopcicHirata2018}.

\subsection{Stellar Spectra}

\begin{figure}
\centering
\includegraphics[width=0.5\textwidth]{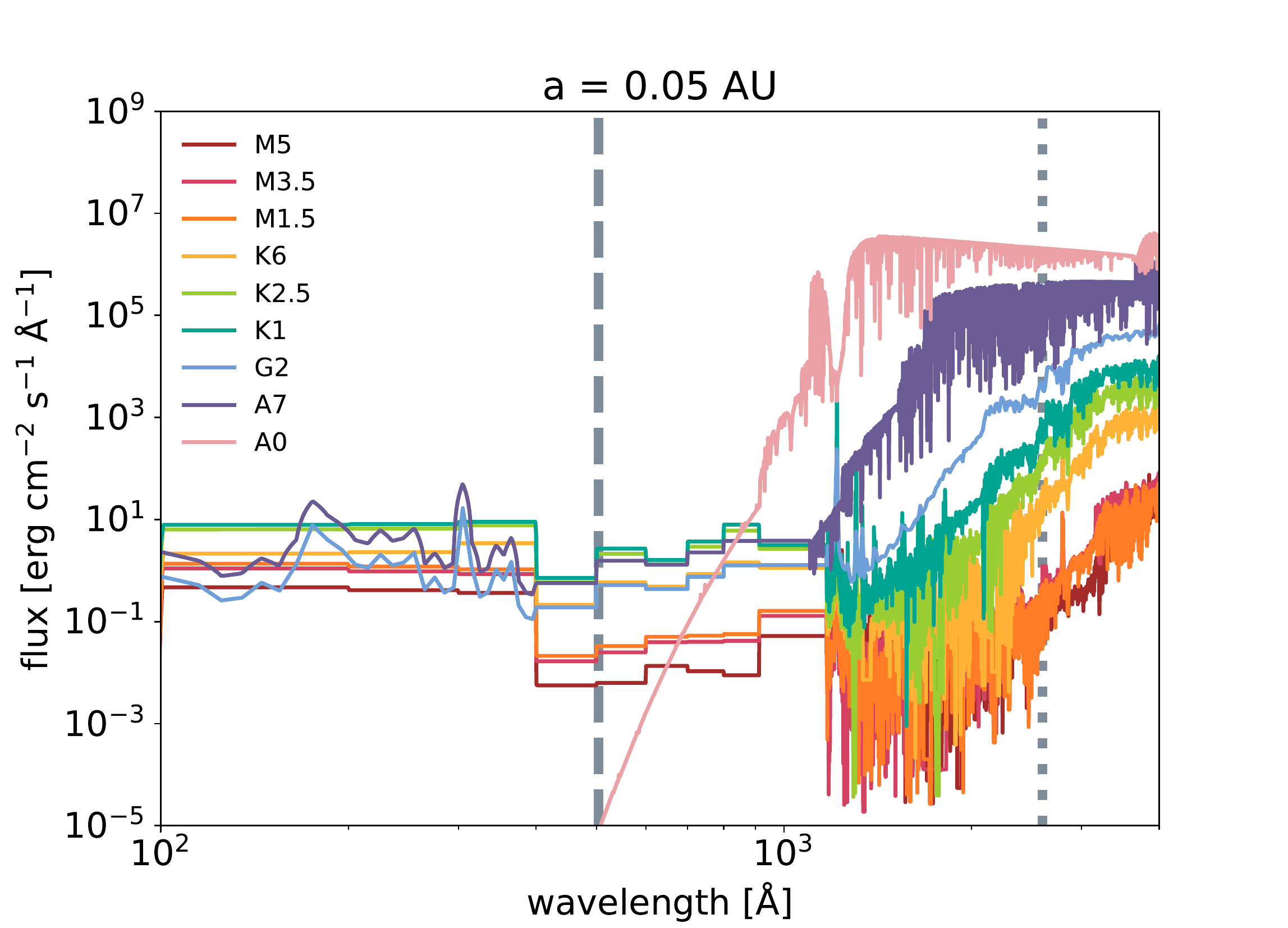}
\caption{Flux density at 0.05~AU from stars of different spectral types. The dashed vertical line shows the ground-state helium ionization threshold (i.e. photons that populate the metastable state) and the dotted line shows the ionization threshold for the metastable triplet state (i.e. photons that depopulate the metastable state).}
\label{fig:stellar_seds_fixeddist}
\end{figure}

Photoionization of hydrogen and helium atoms plays a key role in controlling the triplet helium level population. The corresponding  photoionization rates depend on the shape and intensity of incident stellar radiation field, particularly at wavelengths shortward of those associated with hydrogen ionization ($\lambda = 912$~\AA), the ionization of the helium ground state ($\lambda = 504$~\AA), and the ionization of the excited 2$^3$S helium state ($\lambda = 2600 $~\AA).

In our analysis, we use stellar SEDs for main-sequence stars of different spectral types (see \autoref{fig:stellar_seds_fixeddist}) from the following sources: 
\begin{enumerate}
    \item spectra of late-type stars (K1 to M5) from the MUSCLES survey \citep{France2016,Youngblood2016,Loyd2016}, versions 2.1 (GJ 876, GJ 436, HD 85512) and 2.2 (HD 97658, GJ 667C, HD 40307);
    \item for G2 type, we use the SORCE solar spectral irradiance data from the LASP Interactive Solar Irradiance Data Center,\footnote{\url{http://lasp.colorado.edu/lisird/}} along with scaling relations between the Ly$\alpha$ flux and fluxes in extreme-ultraviolet (EUV) bands from \citet{Linsky2014} to fill the gap in the data between $\sim 400$ and $1150$~\AA;
    \item for A7 and A0 types, we use the synthetic spectra for stars of effective temperatures of 7500 K and 10000 K from \citet{Fossati2018}. The high-energy end of the spectrum for the A7 spectral type was constructed by taking the solar spectrum and multiplying it by a factor of 3.
\end{enumerate}

\section{Results and Discussion}
\label{sec:results}

\subsection{Higher Fraction of Metastable Triplet Helium around Late-type and Active Stars}
\label{sec:stars_fixed_distance}

\begin{figure}
\centering
\includegraphics[width=0.5\textwidth]{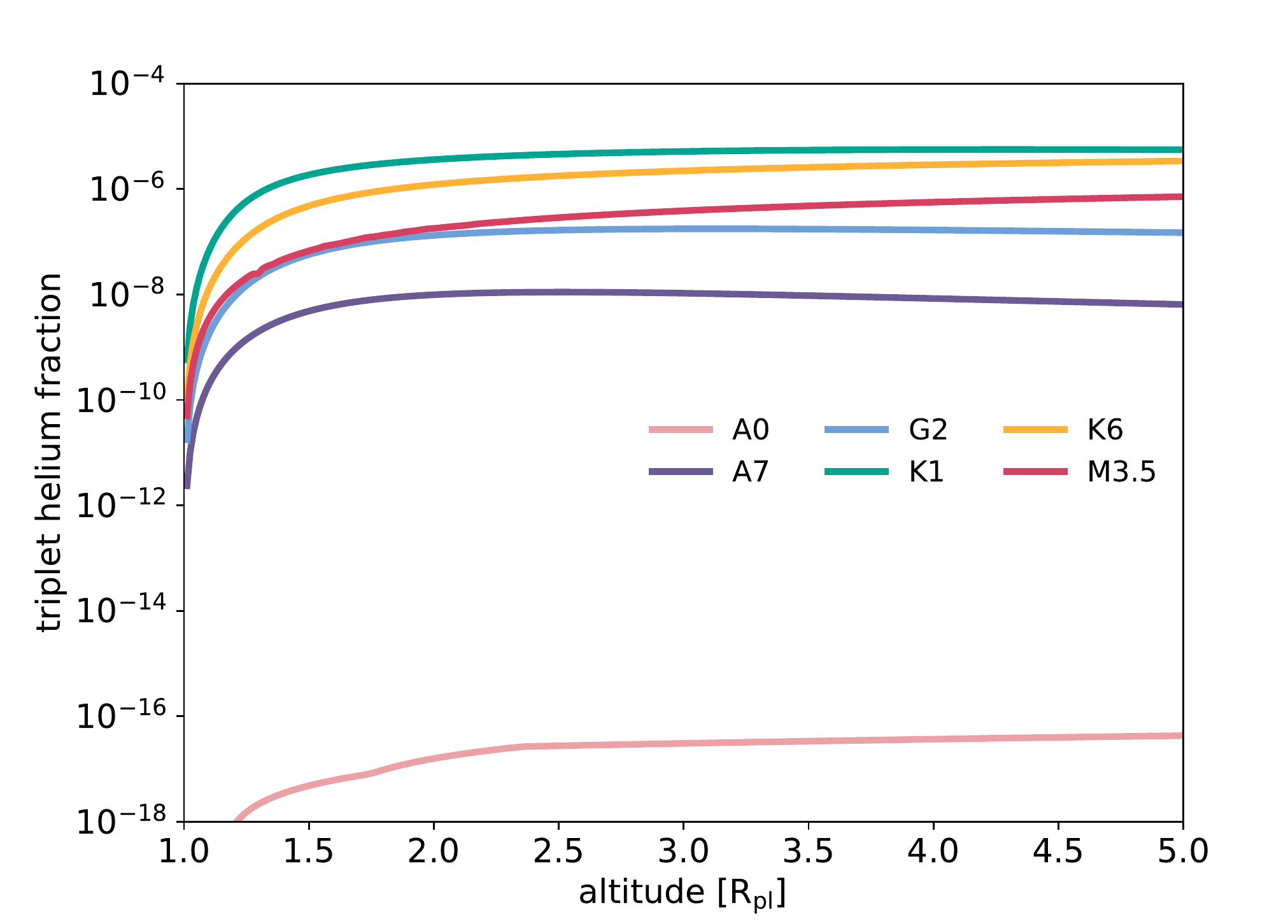}
\caption{Fraction of helium (relative to all He atoms/ions) in a planetary upper atmosphere that populates the excited, metastable triplet state, shown as a function of altitude. Different lines represent (otherwise identical) planetary atmospheres irradiated with stellar SEDs of different spectral types. Higher fraction of triplet helium in a planetary atmosphere produces stronger transit absorption in the 1083 nm line.}
\label{fig:triplet_helium_fraction}
\end{figure}

In order to isolate the effects of incident stellar radiation on the expected 1083 nm absorption from an extended planetary atmosphere, we keep all the parameters describing the planet (mass, radius, and orbital separation) and its escaping atmosphere (mass-loss rate, temperature, and composition) fixed.\footnote{In reality, atmospheric escape rate and thermospheric temperature also depend on the intensity and spectral shape of the incident stellar radiation, in a way that has yet to be observationally inferred and/or tested.} We consider a planet with properties similar to HAT-P-11b, one of the first exoplanets with an extended atmosphere detected via the helium 1083 nm line \citep{Allart2018, Mansfield2018}. The planet mass and radius are set to M$_\mathrm{pl}=0.0736$~M$_\mathrm{Jupiter}$ and R$_\mathrm{pl}=0.389$~R$_\mathrm{Jupiter}$, and the orbital distance to the host star is 0.05254~AU \citep{Yee2018}. For the model temperature and mass-loss rate, we choose $T=7300$~K and $\dot{M}=10^{10.6}$~g~s$^{-1}$, which is a combination of parameters that, when irradiated with a K4-type stellar spectrum, produces an absorption signature that is well matched by the 1083 nm observations of HAT-P-11b by \citet{Mansfield2018} and \citet{Allart2018}.

\begin{figure*}
\centering
\includegraphics[width=1.0\textwidth]{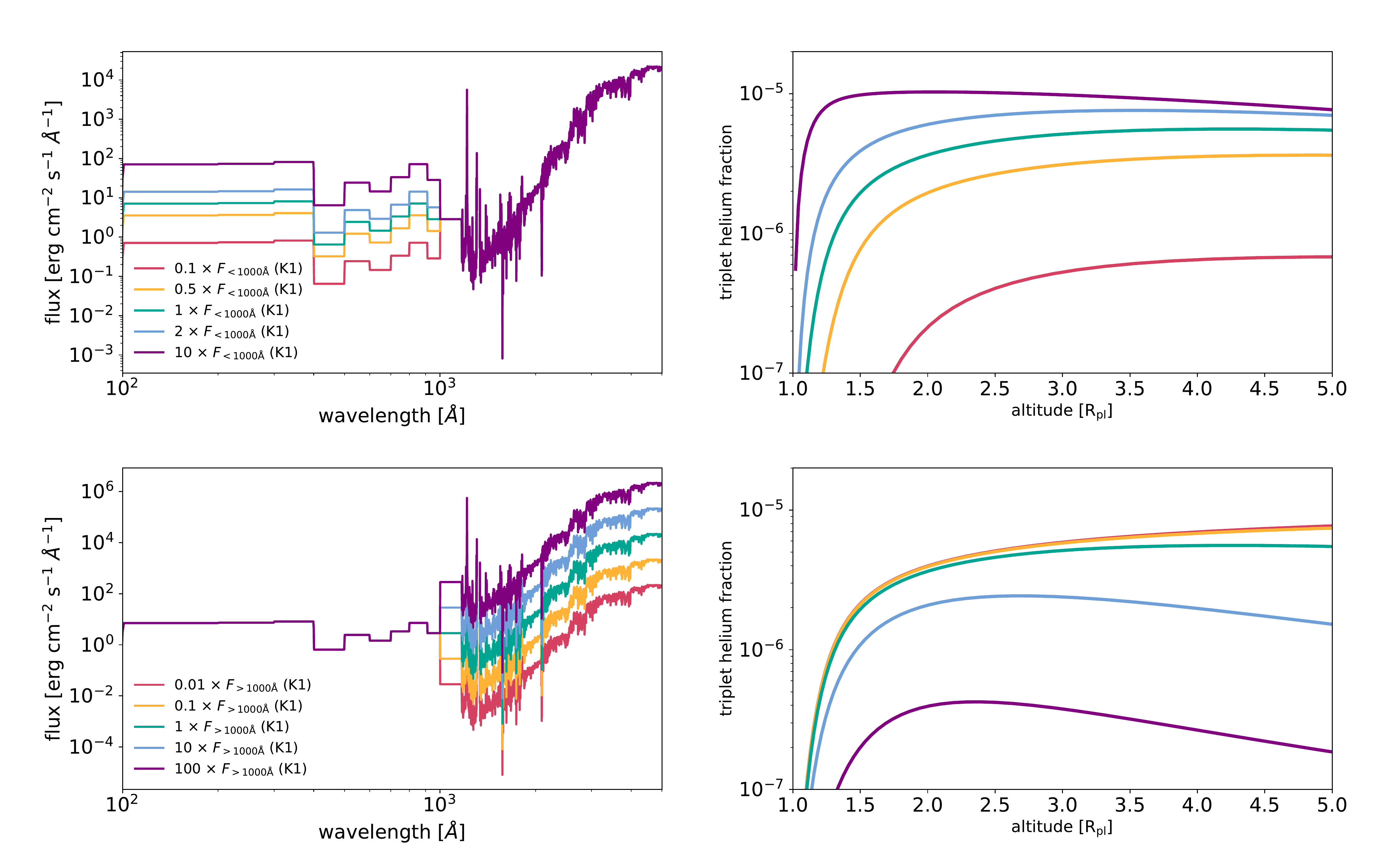}
\caption{Left: green line shows the flux density of a K1 star as seen from a distance of 0.05~AU. Other lines are obtained by multiplying the high-energy end of the spectrum ($\lambda < 1000$~\AA) by a factor between 0.1 and 10 (upper left panel), or the low-energy end of the spectrum ($\lambda > 1000$~\AA) by a factor between 0.01 and 100 (lower left panel). Right: the fraction of helium in the metastable triplet state in a planetary atmosphere irradiated by a stellar spectrum of corresponding color shown on the left. Stellar SEDs with increased levels of high-energy flux and lower levels of low-energy flux produce more helium in the triplet state and consequently, cause stronger absorption in the 1083 nm line.}
\label{fig:K1active}
\end{figure*}

We irradiate the model planetary atmosphere with six different stellar spectra shown in \autoref{fig:stellar_seds_fixeddist} (A0, A7, G2, K1, K6, and M3.5) and for each case we calculate the radial distribution of neutral and ionized hydrogen, as well as the distribution of helium atoms in singlet and triplet configurations. Planetary absorption signal in the 1083 nm line directly depends on the fraction of helium atoms in the triplet state, shown in \autoref{fig:triplet_helium_fraction}. The resulting triplet helium fractions span a wide range of values for different spectral types and show a fairly mild dependence on altitude for any given spectral type. The main conclusion of this analysis is that the triplet helium fraction is higher in late-type stars (K1, K6, and M3.5) than in stars of earlier spectral types (G2, A7, and A0). Although the exact value and the altitude-dependence of the triplet helium fraction depend somewhat on the assumed properties of the planet and its extended atmosphere, the trend shown here---higher triplet fraction in atmospheres around late-type stars---remains unaffected by these changes.

What makes late-type stars particularly favorable for exciting helium atoms in their planets' atmospheres into the metastable triplet state is the hardness of their spectra, i.e. the relative flux intensity at EUV wavelengths, which are responsible for populating the metastable state (dashed line in \autoref{fig:stellar_seds_fixeddist}), and mid-UV wavelengths responsible for depopulating the metastable state (dotted line in \autoref{fig:stellar_seds_fixeddist}). 

In order to demonstrate that both wavelength bands play a role in controlling the population level of excited helium atoms, we take the spectrum of a K1 star and artificially change its high-energy and low-energy part of the spectrum separately. The results are shown in \autoref{fig:K1active}. The upper panels show the effects of changing the high-energy end of the spectrum ($\lambda < 1000$~\AA) by multiplying it by factors of 0.1, 0.5, 2, and 10. As shown in the upper right panel, the higher the EUV and X-ray (i.e. XUV) flux, the higher the population level of metastable triplet helium. Bottom panels show the effects of changing the spectrum at wavelengths $\lambda > 1000$~\AA\ by factors of 0.01, 0.1, 10, and 100. In this case, the population of helium atoms in the triplet state is maximized when the mid-UV flux is minimized. In conclusion, stars with higher levels of XUV flux (due to stellar activity, for example)\footnote{We note that most stellar SEDs obtained in the MUSCLES survey, including the one used in this analysis (of a K1 star HD 97658), come from stars that are not classified as active or flare stars \citep{France2016}.} and lower levels of flux in the mid-UV part of the spectrum (due to lower effective temperature) are most efficient at exciting helium atoms into the triplet metastable state.

\begin{figure}
\centering
\includegraphics[width=0.5\textwidth]{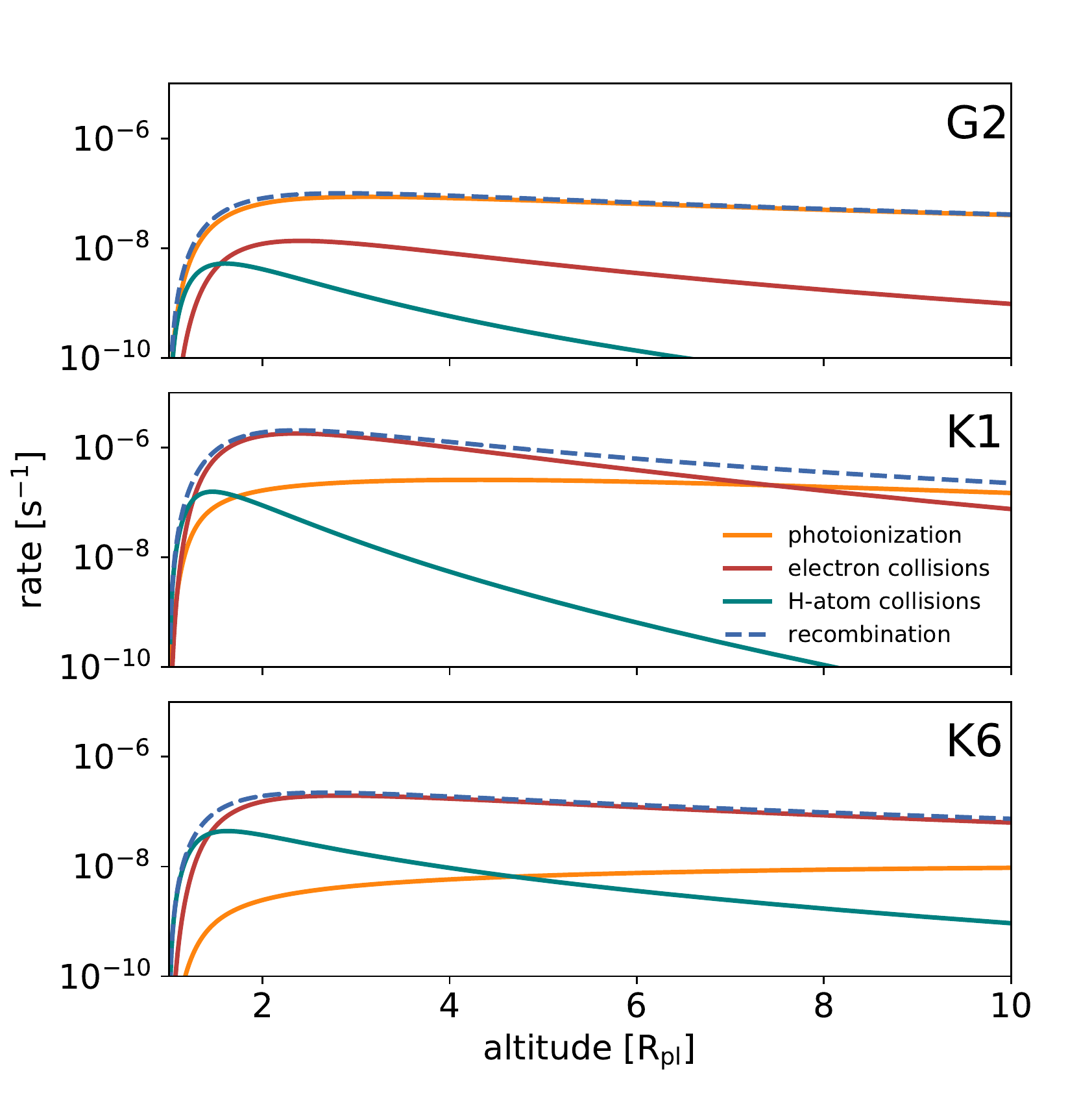}
\caption{Magnitudes of the main reaction rate terms in \autoref{eq:triplet} that control the triplet helium fraction. In G-type and hotter stars, the rate of recombination into the triplet state is balanced by direct metastable-state photoionization. In cooler stars with lower mid-UV flux, the recombination rate is balanced by the rate of collisions that change the configuration of helium atoms from triplet to singlet.}
\label{fig:reaction_rates}
\end{figure}

To provide a physical explanation for this trend, we look at the contribution of each individual term in the helium triplet balance equation (\autoref{eq:triplet}). For the sake of clarity, in \autoref{fig:reaction_rates} we only show the magnitudes of the most significant terms: recombination into the triplet state (first term on the right side of \autoref{eq:triplet}), photoionization of the metastable state (third term), depopulation of the metastable triplet helium state via collisions with electrons (sum of the fifth and sixth term), and collisions with hydrogen atoms (seventh term). Recombination is the main populating mechanism of triplet helium, while the remaining processes shown in \autoref{fig:reaction_rates} depopulate the metastable triplet helium state.

In sufficiently dense planetary atmospheres irradiated by hard spectra, such as those of K-type and cooler stars, helium recombination into the triplet state is balanced by collision-induced triplet-to-singlet transitions. Using the temperature dependence of the rate of helium recombination into the triplet state, $\alpha_3 \approx 2\times 10^{-13} (10^4/T)^{0.8}$~cm$^3$~s$^{-1}$ \citep{OsterbrockFerland}, \autoref{eq:triplet} can be simplified in the following way:
\begin{eqnarray}
\nonumber    f_3 &\approx& \frac{(1-f_1)\alpha_3}{q_{31a}+q_{31b}} = 7\times 10^{-6}\left( \frac{10^4 K}{T}\right)^{0.8}(1-f_1)\\
    &\approx& 7\times 10^{-6}\left( \frac{10^4 K}{T}\right)^{0.8} \left(1-e^{-\Phi_1 e^{-\tau_1}r/v} \right) \ \mbox{.}
    \label{eq:late_stars_approx}
\end{eqnarray}
For systems in this regime, the triplet helium fraction is almost insensitive to variations in the mid-UV part of the spectrum and increases with increasing EUV flux, which controls the ground-state photoionization rate ($\Phi_1$). This dependence is confirmed by the triplet helium fractions in atmospheres irradiated by K1 or harder spectra shown in \autoref{fig:K1active}: green, yellow, and red lines in the bottom right panel practically overlap each other because lowering the mid-UV flux below the K1 level has almost no effect once the system reaches the regime described by \autoref{eq:late_stars_approx}.

Planetary atmospheres irradiated by spectra softer than K1, such as those around G- and earlier-type stars, or atmospheres in which the density of free electrons is too low for collisions to balance helium recombination, recombination is balanced by direct photoionization of the metastable state, driven by stellar flux at mid-UV wavelengths. In that case, \autoref{eq:triplet} reduces to:
\begin{eqnarray}
   & & f_3 \approx \frac{(1-f_1)n_e\alpha_3 }{\Phi_3 e^{-\tau_3}} \label{eq:early_stars_approx}\\ 
\nonumber &\approx &  2\times 10^{-13} \left( \frac{10^4 K}{T}\right)^{0.8}  \frac{\left(1-e^{-\Phi_1 e^{-\tau_1}r/v} \right)}{\Phi_3 e^{-\tau_3}} n_e \ [\mbox{cm}^3 \mbox{s}^{-1}] \mbox{.}
\end{eqnarray}
In this regime, the triplet helium fraction increases with increasing EUV flux (which governs the ground-state photoionization rate, $\Phi_1$) and decreases with increasing mid-UV flux (which governs the photoionization rate of the metastable state, $\Phi_3$). Our calculations of $f_3$ for planets irradiated by spectra softer than K1, shown in \autoref{fig:K1active} (the yellow and red lines in top panels and blue and purple lines in bottom panels) confirm these trends.

\subsection{Higher Fraction of Metastable Triplet Helium at Smaller Orbital Separations}
\label{sec:stars_distance}

\begin{figure}
\centering
\includegraphics[width=0.5\textwidth]{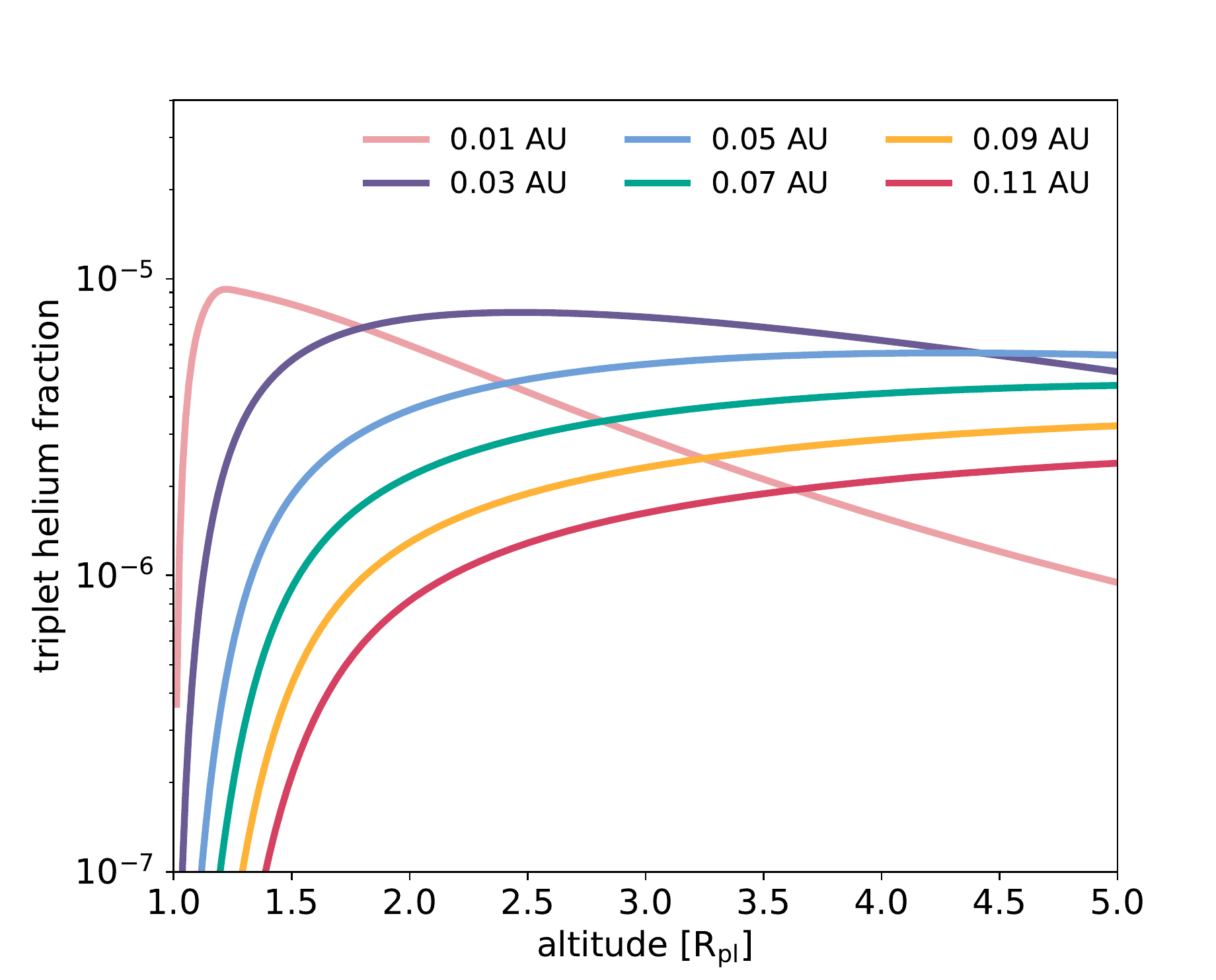}
\caption{Fraction of helium atoms in the metastable triplet state in planetary atmospheres at various orbital distances from a K1 host star. Planets at shorter orbital separations have higher fractions of triplet helium.}
\label{fig:K1distance}
\end{figure}

To investigate how the population of excited helium atoms depends on the magnitude of the received stellar flux, we place our model planetary atmosphere on various orbital distances from a star of K1 spectral type, thereby changing the magnitude of the radiation flux received by the planet. As before, we keep all other stellar and planetary parameters fixed. However, changing the planet--star separation and the level of incident flux could affect the planet's mass-loss rate; we later investigate (separately) how varying $\dot{M}$ changes the helium level population.

In \autoref{fig:K1distance} we show how the resulting fraction of helium atoms in the triplet state depends on the planet's distance from the star, ranging between 0.01 and 0.11~AU. The triplet fraction, especially at lower planetary altitudes, decreases at larger orbital separations. Consequently, we do not expect prominent absorption signals at 1083 nm in exoplanets orbiting main-sequence stars at distances greater than $\sim 0.1$~AU. At very short distances ($\lesssim 0.03$~AU), the fraction of ionized helium in the extended atmosphere increases, thereby reducing the fraction of helium in the (neutral) triplet state at high planetary altitudes. 

This dependence on flux magnitude can help us understand why the triplet helium fraction shown in \autoref{fig:triplet_helium_fraction} is not maximized in planets around stars with the hardest spectra, i.e. M-dwarfs. As shown in \autoref{fig:stellar_seds_fixeddist}, at a fixed orbital distance, the EUV flux of M-stars in the MUSCLES sample is at least an order of magnitude lower than in K-stars. If the EUV flux of M-dwarfs in the MUSCLES sample is representative of high-energy radiation typical for M-stars, then planets irradiated by M-star spectra have to be much closer to their host stars to achieve the same (or higher) population level of excited helium as planets around K-stars. 

\begin{figure}
\centering
\includegraphics[width=0.5\textwidth]{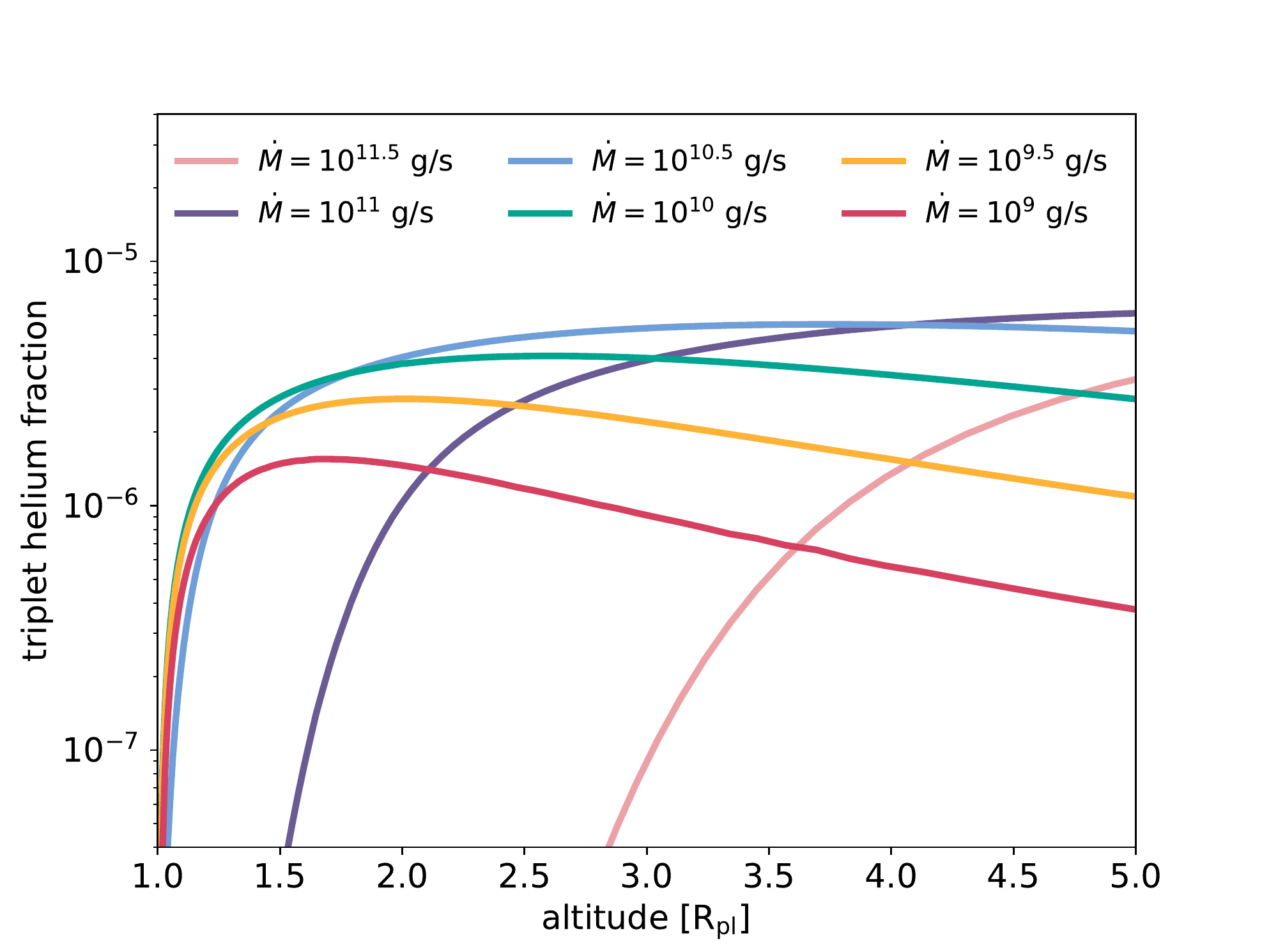}
\caption{Fraction of helium atoms in the metastable triplet state in planetary atmospheres with varying levels of mass-loss rate ($\dot{M}$). For low values of $\dot{M}$, the fraction of triplet helium grows with increasing $\dot{M}$; whereas for high values, the region of the atmosphere closest to the planet becomes depleted of triplet helium due to the lack of free electrons (and the reduced helium recombination rate) caused by hydrogen self-shielding.}
\label{fig:changing_mdot}
\end{figure}

Changing the level of incident stellar flux may, at the same time, affect the basic properties of planetary winds, such as their mass-loss rates. Even though these changes would in reality occur simultaneously, here we treat them separately in order to more easily isolate the effects of each model parameter. In \autoref{fig:changing_mdot} we show the fraction of helium atoms in the metastable triplet state in $T=7300$~K atmospheres at 0.05 AU from a K1 host star, with mass-loss rates varying between $10^9$ and $10^{11.5}$~g~s$^{-1}$. For low mass-loss rates, $\dot{M}\lesssim 10^{10}$~g~s$^{-1}$, the density of the planetary wind (and consequently the density of free electrons) is low and the atmosphere has not reached the regime dominated by electron collisions, described by \autoref{eq:late_stars_approx}; instead, the triplet helium fraction is better described by \autoref{eq:early_stars_approx}. Therefore, as $\dot{M}$ increases, so does the overall atmospheric density, the density of electrons, and the triplet helium fraction. For intermediate values of the mass-loss rate ($\sim 10^{10}-10^{11}$~g~s$^{-1}$), the triplet helium fraction, especially in the part of the atmosphere closest to the planet, is maximized. At high mass-loss rates, $\dot{M}\gtrsim 10^{11}$~g~s$^{-1}$, as the density of the planetary wind increases, the hydrogen atoms begin to self-shield and the inner region of the atmosphere becomes devoid of free electrons. The lack of free electrons reduces the rate of helium recombination, and consequently, the  population of helium atoms in the triplet state close to the planet.

\subsection{Uncertainties Due to the XUV Flux Reconstruction}
\label{sec:EUVreconstruction}

\begin{figure}
\centering
\includegraphics[width=0.5\textwidth]{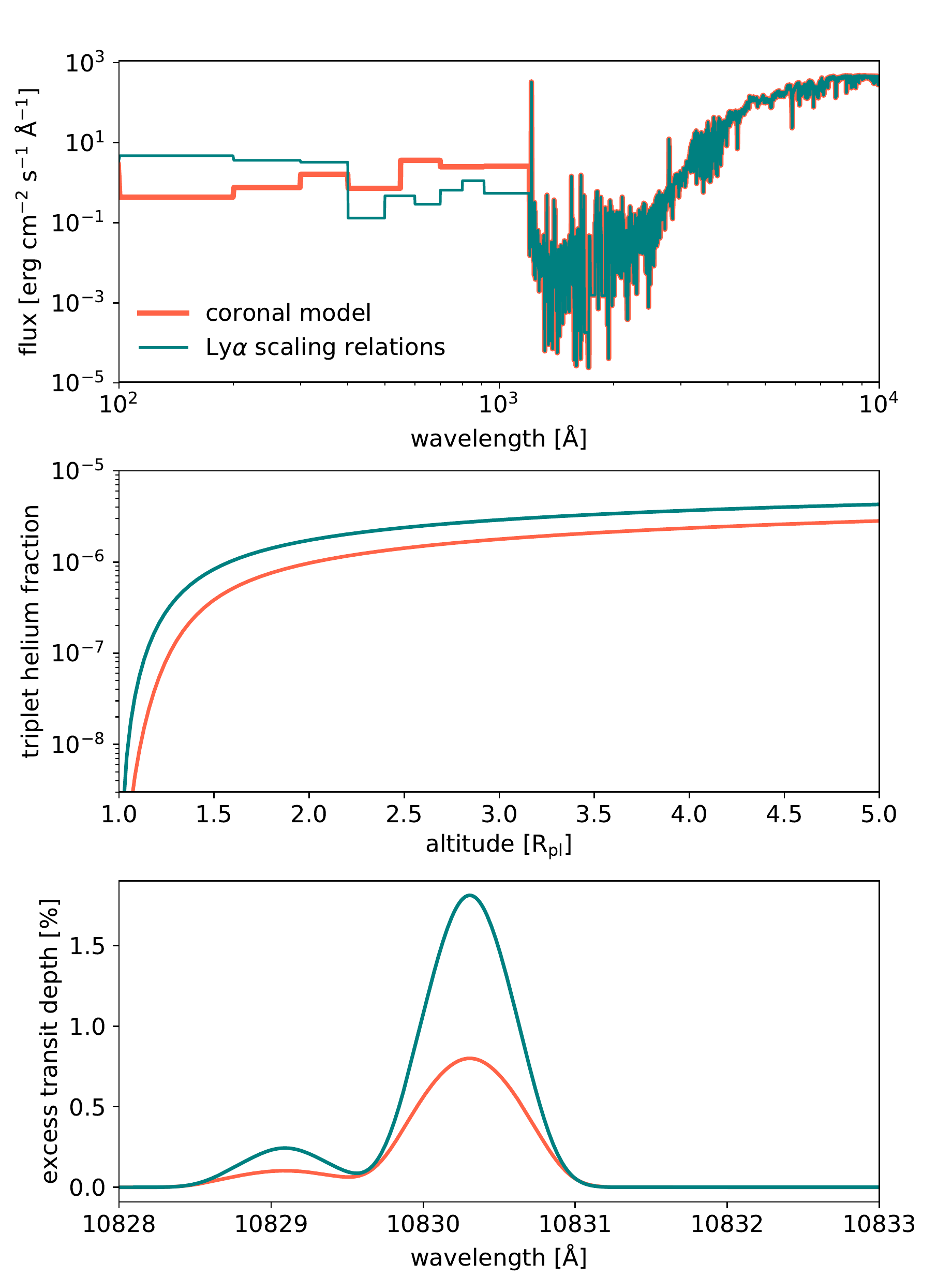}
\caption{Flux density at 0.05 AU from an M1.5-type star constructed from the high-energy ($\lambda <1000$~\AA) end of the spectrum of GJ 3470 from \citet{Bourrier2018} and the low-energy end of GJ 667C from the MUSCLES survey \citep{France2016}. The synthetic high-energy part of the spectrum was constructed using two methods, one based on the X-ray flux and a coronal model (red) and the other based on semiempirical relations from the measured Ly$\alpha$ flux (green). Differences in the stellar EUV flux reconstruction lead to different model predictions for the planetary triplet helium population level (middle panel) and the resulting excess transit depth at 1083 nm (bottom panel).}
\label{fig:M1.5_green_red}
\end{figure}

Stellar spectra, particularly in the EUV and mid-UV wavelength range, play a crucial role in our modeling of helium level populations and the resulting 1083 nm absorption signals from extended exoplanet atmospheres. Direct observations of the EUV flux are currently not possible for stars other than the Sun. Scaling relations and models connecting the EUV flux and other observables, such as X-ray and Ly$\alpha$ fluxes, have been developed \citep[e.g.][]{Sanz-Forcada2011,Linsky2014}. Here we investigate the level of uncertainty caused by the differences between two methods of EUV flux reconstruction.

The red line in \autoref{fig:M1.5_green_red} (top panel) shows the synthetic EUV spectrum of an M-dwarf GJ 3470 calculated using a coronal model based on X-ray and UV spectroscopic data, following \citet{Sanz-Forcada2011}. The green line shows the synthetic EUV spectrum of the same star, obtained using the Ly$\alpha$ flux and semiempirical relations from \citet{Linsky2014}. Both EUV spectra are from \citet{Bourrier2018}. Above $\lambda = 1100$~\AA, we use the spectrum of a different M1.5 star (GJ 667C) from the MUSCLES data.

As in the previous sections, we irradiate a model atmosphere with these two stellar SEDs. For simplicity, we keep all planetary and stellar parameters fixed on values for HAT-P-11b, as described before. In each case, we calculate the population level of triplet helium, shown in the middle panel of \autoref{fig:M1.5_green_red}, and the resulting transit depth at 1083 nm, shown in the bottom panel. Different methods of EUV flux reconstruction can lead to considerable differences in the predicted triplet helium level population and transit depth at 1083 nm. This highlights the importance of obtaining---through direct observations and modeling---reliable stellar spectra at UV wavelengths for stars of various spectral types and levels of activity.

\section{Conclusions}
\label{sec:discussion}

The results of our analysis suggest that the most favorable conditions for maintaining a significant population of helium atoms in the excited 2$^3$S state arise at close orbital separations ($a\lesssim 0.05$~AU) from K-type stars, particularly those with increased levels of activity. Consistent with the theoretical expectation, the first four detections of helium absorption at 1083 nm have been reported for exoplanets that fall into that category (see \autoref{tab:host_stars}). Nondetections have been reported for a few planets around hotter (A- and G-type), as well as cooler (M) stars (see \autoref{tab:host_stars_nondetection}). 

Although M-dwarfs have the hardest spectral shape of all spectral types shown in \autoref{fig:stellar_seds_fixeddist}, their EUV flux level (at a fixed distance from the star) is at least an order of magnitude lower compared to the EUV flux around K-stars. Therefore, planets around M-stars similar to those in the MUSCLES sample must be on much closer orbits in order for their atmospheres to achieve as high fractions of triplet helium as seen around K-stars.

In our analysis, we used the fraction of helium atoms in the triplet state as a proxy for the strength of the absorption signal at 1083 nm. However, it is worth emphasizing that the strength of the absorption signal, i.e. the transit depth at 1083 nm, also depends on the relative size of the observed planet and its host star. Therefore, planets of a given size orbiting around cooler main-sequence stars have a double advantage of having a higher fraction of triplet helium and having a higher planet-to-star radius ratio. 

We do not expect prominent 1083 nm absorption signals from planets orbiting hot, A-type stars. Due to the high level of mid-UV flux around these stars, the population of metastable triplet helium atoms in a planetary atmosphere is easily depleted on short time-scales through direct photoionization.

\acknowledgments
A.O. would like to thank the anonymous referee for their insightful comments, Lynne Hillenbrand and Trevor David for the encouragement to write this paper and for many constructive conversations and helpful comments along the way, Chris Hirata for reading the manuscript, and Luca Fossati for providing SEDs of intermediate-mass stars. A.O. gratefully acknowledges support from an Institute for Theory and Computation (ITC) Fellowship. Support for this work was provided by NASA through the NASA Hubble Fellowship grant \#HST-HF2-51443.001-A awarded by the Space Telescope Science Institute, which is operated by the Association of Universities for Research in Astronomy, Incorporated, under NASA contract NAS5-26555.

\software{matplotlib \citep{Hunter2007}, numpy \citep{numpy}, scipy \citep{Jones2001}.}



\bibliography{refs_helium}



\end{document}